\documentclass{article}
\usepackage{amsfonts, graphicx, stmaryrd, cite}

\def\textbf#1{{\bf #1}}

\def\be{\begin{equation}}
\def\ee{\end{equation}}

\def\ben{\begin{eqnarray}}
\def\een{\end{eqnarray}}
\def\eea{\end{array}}
\def\bea{\begin{array}}

\newcommand{\bei}{\begin{itemize}}
\newcommand{\eei}{\end{itemize}}

\title{Ground State Energy of Mean-field Model of Interacting Bosons in Bernoulli Potential}
\author{Michael Bishop, Jan Wehr}
\date{\today}
\begin{document}
   \maketitle
\begin{abstract}
This paper explores a system of interacting `soft core' bosons in the Gross-Pitaevskii mean-field approximation in a random Bernoulli potential.  First, a condition for delocalization of the ground state wave function is proved which depends on the number of particles and interaction strength.  Using this condition, asymptotics for ground state energy per particle are derived in the large system limit for small values of the coupling constant.  Our methods directly describe the shape of the ground state in a given realization of the random potential.

\end{abstract}
\section{Introduction}

 In his seminal work~\cite{Anderson58}, Anderson discovered that quantum mechanics behave very differently in disordered environments than in periodic environments.  Since this discovery, there has been a lot of effort both in physics and in mathematics  to understand this behavior more completely.  See~\cite{Lewenstein12} which contains a recent physical survey of Anderson localization in a context relevant for the present paper and~\cite{Kirsch07} for an excellent introduction to rigorous mathematical results.\\

The experimental realization of cold atoms~\cite{Anderson1Dexp2, Anderson1Dexp} has added to the interest in the topic and led to new questions about the role of interactions in quantum systems.  Physicists have researched the relation between disorder and interaction~\cite{AAL80, GiamarchiSchulz88, Fisher89, BAA06, SCLBS07}.  There is still much debate on the nature of  phase transitions in both random~\cite{ICFO12} and quasiperiodic potentials~\cite{Roux08}.  Systems similar to the one discussed in this paper have been experimentally realized~\cite{Modugno11}.  For infinite local `hard core' interactions, see~\cite{ALS}.  Lieb, Seiringer, and Yngvason have rigorously shown convergence of  the ground state of a bosonic system with nonrandom harmonic potential to the mean-field approximation ~\cite{LS1, LS2, LS3,LS4}.  The dynamics of the nonlinear Schr\"{o}dinger equation with a random potential is studied in~\cite{Soffer12}; see references therein.\\

Recently, Seiringer, Yngvason, and Zagrebnov considered Bose-Einstein Condensation in systems with randomly placed point scatterers in the continuous setting; they demonstrated the existence of a condensate in such systems (under certain conditions on the interaction) as well as a presence of a phase transition.~\cite{SYZ12}.  Their analysis is based on a detailed study of the statistics of random potential realizations, similar to what we do here.  Our model can be viewed as a discrete version of the model considered in~\cite{SYZ12}, with the space variable rescaled by $L$ and the parameters in the Hamiltonian and the random potential scaling as follows:  $\gamma = \gamma_0 L^2,\ \sigma = \sigma_0 L,\ \nu = \nu_0 L.$ \\

For interacting systems in random potentials, Anderson localization competes with delocalization caused by repulsive interaction. It is not obvious which of the two mechanisms dominates the system's behavior, even in the ground state.  One approach to this question is to start from a noninteracting system and treat interaction as a perturbation.  A system of bosons with no interaction places each particle in the single particle ground state.  When interactions are added to a finite-volume system, they can be controlled for small values of $g$. \\

The goal of this paper is to study effects of disorder in a system of bosons which interact with a weak repulsive `soft core' force as the number of particles and system size are taken to infinity, proving two results.  Theorem 1 is a general statement about delocalization effects of such interaction, applying to both discrete and continuous versions of the model.   It is then applied, together with detailed analysis of the energy functional, to the one-dimensional system with a Bernoulli-distributed  potential---Theorems 2 and 3.  We derive there an asymptotic formula for the ground state energy per particle as the product of the interaction strength and the particle density goes to zero.  We directly study the way the geometry of each realization of the random potential determines the ground state wave function.  The methods are inspired by an adaptation of the technique used in ~\cite{BishopWehr12}.\\

In the discrete setting, the state of $N+1$ bosons in the lattice cube $\Lambda = \{0, \dots, L+1 \}^d$ with length $L$ is described by a normalized wave function $|\Phi(x_1, \dots, x_{N+1})\rangle$, a function in the symmetric subspace of $\otimes^{N+1} L^2(\Lambda, \mu)$ with Dirichlet boundary conditions, where the measure $\mu$ is the counting measure  and $x_i$ are positions of the particles in $\Lambda$.  The Hamiltonian of the system is given by
\be H = \sum_{i=1}^{N+1} H_i + \sum_{i\neq j} U(x_i , x_j) \ee 
where $H_i $ is the single particle Hamiltonian acting on the $i$-th particle and $U$ is the potential of the interaction between particles.  The single particle Hamiltonian is the (random) Schr\"{o}dinger operator 
\be H = -\Delta + V\ee
where $\Delta$ is the discrete Laplacian and $V$ is a (random) multiplication operator, which in this paper will always be bounded below (without loss of generality by $0$).  The interaction $U$ is a `soft core' interaction of the form
\be g\delta_{x_i, x_j}\ee
where the $\delta$ is the Kronecker delta and the coupling constant $g$ is positive, making the interaction repulsive.  This repulsion makes it energetically unfavorable for bosons to occupy the same space, but does not exclude the possibility, unlike in the the case of 'hard core' interactions where $g=\infty$.  Because of the difficulty of working in a large tensor space, the multi-particle bosonic states are approximated by the Gross-Pitaevskii mean-field wave function~\cite{Annett}.  In this approximation, each boson is assumed to be in the same state $\phi$ which defines the state of the whole system:  $|\Phi\rangle = |\phi(x_1)\rangle\dots|\phi(x_{N+1})\rangle$.  The Hamiltonian becomes 
\be (N+1) H_i + \frac{g(N+1)(N)}{2}|\phi|^2, \ee 
a nonlinear random Schr\"{o}dinger operator (the second term acts as a potential, which depends on the state itself---hence the nonlinearity).  The approximation exchanges the difficulties arising from Bose statistics for the nonlinearity of the new problem.  The associated per particle energy functional is 
\be E[\Phi] =  \sum_{x\in\Lambda} \sum_{|y| =1} |\phi(x+y) - \phi(x)|^2 + V(x)|\phi(x)|^2 + \frac{gN}{2}|\phi(x)|^4\ee

{\bf Remark}:  In the infinite volume limit, if the particle number increases proportionally to the system size, the total energy diverges and the natural quantity is energy per particle. In this paper, any mention of energy refers to the per particle energy.

In the continuous setting, the state of $N+1$ bosons in the box $\Omega = [0, L]^d$ with linear size $L$ is described by the wave function $|\Phi(x_1, \dots, x_{N+1})\rangle$, a function in the symmetric subspace of $\otimes^{N+1} L^2(\Omega, \mu)$ with Dirichlet boundary conditions, where the measure $\mu$ is the Lebesgue measure, and $x_i$ are the positions of the particles in $\Omega$.  The Hamiltonian of the system is given by
\be H = \sum_{i=1}^{N+1} H_i + \sum_{i\neq j} U(x_i , x_j) \ee 
where $H_i $ is the single particle Hamiltonian acting on the $i$-th particle and $U$ is the interaction potential between particles.  The single particle Hamiltonian is given by the (random) Schr\"{o}dinger operator 
\be H = -\Delta + V_i\ee
where $\Delta$ is the Laplacian and $V$ is a (random) multiplication operator.  The potential is bounded below (without loss of generality by $0$).  The interaction $U$ is a `soft core' interaction of the form
\be g\delta(x_i-x_j)\ee
where the $\delta$ is the Dirac delta and $g >0$ corresponds to a repulsive interaction.  In the mean-field aprroximation, each boson is assumed to be in the same state $\phi$ which defines a state of the whole system:  $|\Phi\rangle = |\phi(x_1)\rangle\dots|\phi(x_{N+1})\rangle$.  The Hamiltonian becomes 
\be (N+1) H_i + \frac{g(N+1)(N)}{2}|\phi|^2, \ee 
a nonlinear Schr\"{o}dinger operator.  The associated per particle energy functional is 
\be E[\Phi] = \int_\Omega dx \left[|\phi'(x)|^2 + V(x)|\phi(x)|^2 + \frac{gN}{2}|\phi(x)|^4\right]\ee

In section 2, we address the localization and delocalization of states in this system for both the discrete and continuous settings.  For any state $\phi$, define the set
\be X_{>\epsilon}(\phi) = \{x: |\phi(x)| > \frac{\epsilon}{L^{d/2}}\} \ee 
as the set of points where the absolute value of $\phi(x)$ is greater than its average magnitude per site, multiplied by a constant $\epsilon$.  This is a natural set in the study of localization of low energy states. 

{\bf Theorem 1}: Assume that $V \geq 0$, for both the discrete and continuous settings, for a state $\phi$ with energy $E' = E[\phi]$, the measure of the set $X_{> \epsilon}$ obeys the lower bound
\be \mu(X_{>\epsilon}(\phi)) \geq \frac{gN(1-\epsilon^2)^2}{2E'} \ee

This theorem implies that Anderson-type localization for low energy states, $E' \approx 0$, requires strong conditions on $gN$.   For a large number of particles or strong interaction, the wave function must fill a significant amount of space, meaning the repulsion dominates the Anderson localization effects caused by the random potential $V$.  The repulsive interaction, though local, forces overlap to become energetically expensive if too many bosons occupy the same place, causing the ground state to spread.  If a low energy state is localized to some length $\ell$, $gN$ must be smaller than $\ell^d$.  To recover an Anderson type localization, a necessary condition is $gN \leq O(E' \ell^{d})$.
In physical experiments, the interaction constant $g$ is a controlled parameter rather than a variable dependent on the particle number and the particle density is approximately constant ($N \approx \rho L^d$) to ensure the existence of thermodynamic limits.  In these systems, a low-energy state must occupy a nonzero fraction of the volume rather than being localized with some lower order localization length $\ell$.  




In section 3, the above theorem is used to describe the ground state of the one-dimensional lattice system where the potential $V$ is Bernoulli-distributed, the interaction constant $g$ is small, and the particle density is approximately constant: $N+1 \approx \rho L$.  The ground state minimizes the energy (per particle) given by
\be E(\phi) = \sum_{x=1}^L \left(|\phi'(x)|^2 + V(x)|\phi(x)|^2 + \frac{g\rho L}{2}|\phi(x)|^4 \right) \ee
where $\phi \in \ell^2\{0, \dots, L+1\}$ with Dirichlet boundary conditions and $V$ is a multiplication operator by a function $V(x)$ (of a discrete argument), where $V(x)$ are IID Bernoulli random variables: $P[V(x) = 0] = p$, $P[V(x) = b] = 1 - p  = q$, and $b>0$ is a constant.  \\

{\bf Definition}: The {\bf finite volume ground state} is defined as the normalized state $\phi^{(L)}_0$ which minimizes the energy functional $E[\phi]$ and the corresponding energy, $E^{(L)}_0$, is called the {\bf ground state energy}.  \\

{\bf Theorem 2}:  For any $g$, $\rho$, $p$, and $b$,
\be \lim_{L \to \infty} E_0^{(L)} = E_0 \ee
where $E_0$ is a nonrandom function of the above parameters.\\ 

After taking the infinite volume limit, we want to understand the behavior of the ground state for small $g\rho$, when the nonlinear term is taken to zero.  \\

{\bf Theorem 3}: For the one-dimensional lattice Gross-Pitaevskii model with Bernoulli disorder, the ground state energy $E_0$ satisfies the following condition with probability one.

\be 0 < \liminf_{g\rho \to 0} E_0 \log_p^2(g\rho) \leq \limsup_{g\rho \to 0} E_0  \log_p^2(g\rho) < \infty \ee

Theorem 3 is an illustration of Theorem 1.  In Theorem 3, the ground state is approximated by sine waves on intervals of zero potential longer than some minimum interval length and by zero everywhere else.  The ground state energy is bounded above by $\frac{C_+}{(\log_p(g\rho))^2}$.  Using Theorem 1, $\mu(X_{>\epsilon}(\phi))$ is bounded below by 
\be \frac{g\rho L(1-\epsilon^2)^2(\log_p(g\rho))^2}{2C_+} \ee
which is proportional to the system size.  The proof of Theorem 3 shows that for the ground state this lower bound is asymptotically accurate.  \\

{\bf Remark}: We expect that the upper and lower limit in the above statement are equal.  One should be able to close the gap between our lower and upper bounds using  more accurate variational functions---the solution of the discrete nonlinear Schr\"{o}dinger equation on an interval with Dirichlet boundary conditions, which can be thought of as a discrete version of the Jacobi elliptic sine function.   \\

\newpage

\section{Proof of Theorem 1}

{\bf Theorem 1}: If the energy of a state is $E' = E[\phi]$ and the potential $V$ is nonnegative, then for $\epsilon \in (0,1)$ 

\be  \mu(X_{>\epsilon}(\phi)) \geq \frac{gN(1-\epsilon^2)^2}{2E'} \ee
{\bf Proof}: The $\ell^2$-norm of the state $\phi$ restricted to $X_{\leq \epsilon}$ is bounded by
$$\|\phi(x)|_{X_{\leq \epsilon}}\|^2 \leq \mu(X_{\leq \epsilon})  \frac{\epsilon^2}{L^{d}} \leq \epsilon^2$$
Because the state $\phi$ has $L^2$-norm equal to one,  
$$\|\phi(x)|_{X_{> \epsilon}}\| \geq 1 - \epsilon^2$$
The energy of $\phi$ is bounded below by its interaction energy on $X_{> \epsilon}$.  The interaction energy is bounded below using Schwarz's Inequality:
$$\left(\int_{X_{> \epsilon}} |f|^2 d\mu \right)^2 \leq \left(\int_{X_{> \epsilon}} d\mu\right) \left(\int_{X_{> \epsilon}} |f|^4 d\mu \right)$$
which bounds the $L^4$-norm below by
$$\int_{X_{>\epsilon}(\phi)} |f|^4 d\mu  \geq \frac{(1-\epsilon^2)^2}{\mu(X_{>\epsilon}(\phi))}$$
The interaction energy is thus bounded below by
$$ \frac{gN(1-\epsilon^2)^2}{2\mu(X_{>\epsilon}(\phi))}$$
and bounded above by $E'$.  It follows that  
$$\frac{gN(1-\epsilon^2)^2}{2\mu(X_{> \epsilon})} \leq E'$$
which gives the desired lower bound:
\be \mu(X_{\geq \epsilon}) \geq \frac{gN(1-\epsilon^2)^2}{2E'} \ee
and completes the proof. \hfill $\Box$\\

{\bf Remark}:  By shifting the energy, one can easily generalize the above theorem to arbitrary potentials bounded below.  If $V \geq V_{min}$, 
$$ \mu(X_{>\epsilon}(\phi)) \geq \frac{gN(1-\epsilon^2)^2}{2(E' - V_{min})}$$\\

\section{Ground State Estimates for Weak Interaction in Bernoulli Potentials}

\indent Theorems 2 and 3 are specifically for the system on the one-dimensional lattice with Bernoulli potential.\\


In the proof of Theorem 2, we will apply Kingman's subadditive ergodic theorem, using the version from the standard probability reference~\cite{Durrett}; Kingman's original formulation~\cite{Kingman68} would also be sufficient for our purposes. \\

{\bf Kingman's Subadditive Ergodic Theorem}:  Suppose $X_{m,n}$, $0\leq m < n$ satisfy:

(i) $X_{0,m} + X_{m,n} \geq X_{0,n}$.

(ii) $\{ X_{nk, (n+1)k},\ n\geq 1\}$ is a stationary sequence for each $k$.

(iii) The distribution of $\{ X_{m, m + k},\ k\geq 1\}$ does not depend on $m$.

(iv) $\mathbb{E}[X^+_{0,1}] < \infty$ and for each $n$, $\mathbb{E}[X_{0, n}] \geq \gamma_0 n$, where $\gamma_0 > - \infty$.\\

Then

(a) $\lim_{n\to\infty}\mathbb{E}[X_{0, n}]/n = \inf_m \mathbb{E}[X_{0, m}]/m \equiv \gamma$.

(b) $X = \lim_{n\to\infty} X_0 /n $ exists almost surely and in $L^1$, so $\mathbb{E}[X] = \gamma$.

(c)If all the stationary sequences in (ii) are ergodic then $X = \gamma$ almost surely.\\

{\bf Proof of Theorem 2}:  We have
\be E_0^{(L)} = \min _{\|\psi\| = 1} \sum_{j=0}^{L-1} |\psi(j+1) - \psi(j)|^2 + V(j)|\psi(j)|^2 + \frac{g\rho L |\psi(j)|^4}{2}\ee
where $\psi \in \ell^2\{0, \dots L\}$ with Dirichlet boundary conditions. With $\phi = \sqrt{L}\psi$, we can rewrite $E_0^{(L)}$ as
\be E_0^{(L)} = \frac{1}{L}\min _{\|\phi\| = \sqrt{L}} \sum_{j=0}^{L-1} |\phi(j+1) - \phi(j)|^2 + V(j)|\phi(j)|^2 + \frac{g\rho |\phi(j)|^4}{2}\ee
which is equal to $\frac{1}{L} X_{0,L}$, where   
\be X_{0,L} = \min _{\|\phi\| = \sqrt{L}} \sum_{j=0}^{L-1} |\phi(j+1) - \phi(j)|^2 + V(j)|\phi(j)|^2 + \frac{g\rho |\phi(j)|^4}{2}\ee 
The process $X_{0,L}$ satisfies the assumptions of Kingman's theorem.  For (i), $X_{0,L}$  is subadditive since the only difference between $X_{0,M} + X_{M, L}$ and $ X_{0,L}$ is the restriction $\phi(M) = 0$ in the definition of the former.  Properties (ii) and (iii) hold because the $V(j)$ are independent.  For (iv), $X^+_{0,1} < 2 + b + \frac{g\rho}{2}$ and $X_{0, n} \geq 0$.  \\
Therefore, 
\be \lim_{L \to \infty}  E_0^{(L)} = \lim_{L\to\infty}\frac{X_{0,L}}{L} = E_0\ee
almost surely.\hfill$\Box$\\

Theorem 3 says, in essence, that the limit of $E^{(L)}_0$ as $L \to \infty$ to leading order is $\frac{C}{(\log_p(g\rho))^2}$ for some constant $C$ when $g\rho$ small.  The proof of Theorem 3 does give the relationship for small $g\rho$, however it does not provide the precise value of $C$.  However, the proof of Theorem 3 does provide a good approximation of the true ground state.\\

{\bf Theorem 3}: For the one-dimensional lattice Gross-Pitaevskii model with Bernoulli disorder, the ground state energy $E_0$ satisfies the following condition with probability one.

\be 0 < \liminf_{g\rho \to 0} E_0 \log_p^2(g\rho) \leq \limsup_{g\rho \to0} E_0  \log_p^2(g\rho) < \infty \ee



The structure of the ground state depends on the distribution of the intervals of zero potential.  These intervals are colloquially referred to as `lakes' or `islands.'  A realization of the potential $V$ is determined by alternating intervals of zero potential and positive potential.  In a sequence of Bernoulli random variables, the lengths of intervals of zero potential are independent and distributed geometrically: $P[L_i = x] = qp^{x}$.  If the system size $L$ is fixed, the intervals of zero and positive potential are not independent; they are subject to the condition that the sum of their lengths is exactly $L$.  Their number is not constant, but it easy to show that it satisfies a law of large numbers.  The considered system can by approximated by a system of variable length, in which the number of intervals is fixed at the value dictated by the law of large numbers.  Such an approximation was carried out in detail in~\cite{BishopWehr12} by standard probabilistic methods and will not be repeated here.  \\

Let us thus fix the number of independent intervals of zero and positive potential to be $2n$.  This makes $L = \sum_{i=1}^n (L_i + \tilde{L_i})$ a random variable, the sum of $n$ geometrically distributed intervals of zero potential and $n$ geometrically distributed intervals of positive potential .  There intervals of zero potential have lengths $L_i$ with the distribution
\be P[L_i = x] = qp^{x-1}\ee
for integer values of $x$ greater than or equal to one (an interval must have at least one site of zero potential). Likewise, lengths of there intervals of positive potential are distributed according to
\be P[\tilde{L_i} = x] = pq^{x-1} \ee
The variable $i$ indexes the intervals and takes values $1, \dots, n$.  The total system size $L$ is the sum of the random interval lengths.  The system size $L$ has expected value 
\be \mathbb{E}\left[\sum_i^n (L_i + \tilde{L_i}) \right] = \frac{n}{pq}\ee 

{\bf Remark}: By the Law of Large Numbers~\cite{Durrett}, with probability one in the limit $n\to\infty$, the difference of both $L$ and $\sum_{L_i > x} L_i$ and their expectations has order less than $n$. 
These controls occur with probability one in the limit $n \to \infty$, further referred to as:
\be |L - \frac{n}{pq}| = o(n) \ee
\be |\sum_{L_i \geq x} L_i - \mathbb{E}[\sum_{L_i \geq x} L_i]| = o(n) \ee 
where 
$$ \mathbb{E}\left[\sum_{L_i > x} L_i\right] = n\sum_{y \geq \lfloor x\rfloor + 1} y P\left[L_i = y \right]$$
\be = \frac{n}{pq}\left( \lfloor x\rfloor qp^{\lfloor x\rfloor + 1} + p^{\lfloor x\rfloor+1}\right)\ee\\

The result of Theorem 3 is obtained using the following strategy.  An upper bound on the ground state energy can be generated by evaluating the energy functional on any test function.  The method is to find a test function with asymptotics similar to a demonstrable lower bound.  The process is iterative; first, a test function is evaluated to find an upper bound on the ground state energy.  Second, this upper bound is used on the norm of the true ground state restricted to various sets, such as the set of sites of positive potential, in order to isolate the sites where the $\ell^2$-norm of the ground state is concentrated.  Third, the energy of the ground state on these sets is minimized to find a lower bound.  This lower bound may give intuition for a better choice of test function for the upper bound.  The process repeats until the upper and lower bounds on the ground state energy are asymptotically similar. The proof of Theorem 3 is simply the final iteration of this process.\\

{\bf Proof of Theorem 3, Upper Bound}:
The energy of any test function bounds the ground state energy from above.  Consider the test function $\psi$ defined as follows: for an interval of zero potential with length $L_i$, the function is a sine wave $m_i\sqrt{\frac{2}{L_i+1}}\sin(\frac{\pi x}{L_i+1})$, where $m_i$ is the $L^2$-norm of the function restricted to the interval.  On intervals of high potential, $\psi$ is zero.  For intervals of zero potential with length $L_i > \log_p(g\rho) + \log_p\left(\log_p(g\rho)\right)$, we let   
\be m_i^2 = \frac{L_i}{\sum_{L_i > \log_p(g\rho) + \log_p\left(\log_p(g\rho)\right)} L_i} \ee
and for shorter intervals, $m_i = 0$.  This makes the kinetic energy term and interaction energy term have the same asymptotic order as $g\rho \to 0$.  This test function also satisfies the normalization criterion $\sum_i m_i^2 =1$.  The kinetic energy of the discrete sine wave on a specific interval of zero potential is bounded above by $\frac{m_i^2\pi^2}{(L_i+1)^2}$~\cite{BishopWehr12}. The interaction energy, $\frac{g\rho L}{2} \sum_x \|\phi(x)\|^4$ is equal to $\frac{3g\rho L m_i^4}{4L_i}$.  Summing the upper bound on kinetic energy and the interaction energy of test function $\psi$ over the space, the total energy of the test function $\psi$ is bounded above by  
\be \frac{3g\rho L}{4 \sum_{L_i > \log_p(g\rho) + \log_p\left(\log_p(g\rho)\right)} L_i } + \frac{\pi^2}{\left( \log_p(g\rho) + \log_p\left(\log_p(g\rho)\right)+1\right)^2}\ee
where the second term is an overestimate of the kinetic energy, treating each interval as the shortest interval admitted.  Both $L$ and $\sum_{L_i > \log_p(g\rho) + \log_p\left(\log_p(g\rho)\right)} L_i$ depend on the realization of the potential, but by equations (27-29), with probability one in the limit $n \to \infty$,
\be L = \frac{n}{pq} + o(n) \ee
and using $\lfloor x \rfloor \geq x - 1$,
$$\sum_{L_i > \log_p(g\rho) + \log_p\left(\log_p(g\rho)\right)} L_i$$ 
\be \geq \frac{n}{pq} g\rho \log_p(g\rho) \left(p +  q\log_p(g\rho) + q\log_p\left(\log_p(g\rho)\right)\right) + o(n) \ee
by equation (31).  The interaction energy is bounded in the limit as $n \to \infty$ with probability one by
$$\frac{3 }{ 4q \log_p(g\rho)\left[\log_p(g\rho) + \log_p\left(\log_p(g\rho)\right) + \frac{p}{q}\right]}$$
In the limit  as $g\rho \to 0$, the leading order term of the upper bound on the ground state energy is
$$\frac{C'}{\log^2_p(g\rho)}$$
with $C' = \frac{3}{4q} + \pi^2$.  Thus, $\limsup_{g\rho \to \infty} E_0  \log_p^2(g\rho) < \infty$. \hfill $\Box$\\


{\bf Proof of Lower Bound}:

For each $n$ and $g\rho$, the ground state $\phi_0$ is well-defined but not explicitly known.  The lattice will be partitioned into four sets: the set of sites of high potential, denoted $M_b$; the set of sites on intervals of zero potential longer than $\log_p(g\rho)$, denoted $M_{long}$; the set of sites on intervals of zero potential shorter than $\log_p(g\rho)$ where the kinetic energy cannot be easily bounded below, denoted $M_{light}$; and the set of sites on intervals of zero potential shorter than $\log_p(g\rho)$ where the kinetic energy can be easily bounded below, denoted $M_{heavy}$.  \\

The lower bound is shown as follows.  The ground state energy is bounded below by a lower bound of the ground state energy restricted to $M_{heavy}$.  The kinetic energy for a given interval in $M_{heavy}$  is bounded below by Lemma 1.  The interaction energy is bounded below by the bound of Theorem 1.  Lemma 2 provides a lower bound on the norm of $\phi_0$ restricted to $M_{heavy}$, which converges to one in the limit $g\rho \to 0$.  Using Lagrange multipliers, the lower bound on kinetic and interaction energy on each interval is minimized over the $m_i$, the norms of restrictions of the state to each interval.  This minimization determines a minimal interval length for any interval in $M_{heavy}$.  The number of sites in $M_{heavy}$ is estimated above by the number of sites on intervals longer that this minimal interval length.  Using this and the lower bound of the norm of $\phi_0$ restricted to $M_{heavy}$, we obtain the desired lower bound of the ground state energy.\\

Without loss of generality, the ground state wave function can be assumed to be non-negative.  By a standard argument: if the ground state does not have the same complex phase at each site, the kinetic energy can by reduced by making the phases equal.  Since the potential and interaction energy only depend on the magnitude of the state at each site and not on the complex phase, the energy of the resulting state is strictly smaller.  Therefore the ground state must have the same complex phase and can be assumed to be positive.\\


To define $M_{light}$ and $M_{heavy}$, the ground state energy on an interval of zero potential is approximated by that of the the minimizer of the kinetic energy on the interval---the discrete sine wave.  For a given interval with length $L_i$, the ground state determines boundary values $m_i\delta_i^L$ and $m_i\delta_i^R$ on the sites of high potential to the left and right of the interval, respectively, where $m_i$ is the norm of the ground state on the interval.  The boundary values are positive by the above argument.  The kinetic-energy-minimizing state is of the form 
\be \frac{c_im_i}{\sqrt{L_i+1}}\sin(\frac{s_i\pi x}{L_i+1} + t_i)\ee
where the sine wave is normalized to $m_i$ by $c_i \in [1, \sqrt{2}]$, stretched by $s_i \in (0,1)$, and shifted by $t_i$; all three are determined by the $\delta_i$'s and $m_i$~\cite{BishopWehr12}.  Heavy intervals have relatively small $\delta_i$'s which determine a lower bound on kinetic energy.  Light intervals have large $\delta_i$'s which do not admit a good lower bound on kinetic energy, but do allow an upper bound on the norm of the ground state on this intervals.\\

{\bf Definition}:  For intervals of zero potential with length less than $\log_p(g\rho)$, an interval is in $M_{heavy}$ if for the ground state $\phi_0$,
\be \max(\delta_i^L, \delta_i^R) \leq \frac{1}{2\sqrt{L_i}} \ee
An interval is in $M_{light}$,
\be \max(\delta_i^L, \delta_i^R) > \frac{1}{2\sqrt{L_i}} \ee
These definitions can be restated using $m_i$ and are thus directly determined by the  values of the ground state wave function on the sites adjacent to the zero potential intervals.  The definition is stated above without $m_i$ in order to separate the shape and curvature of the sine wave from its norm and to facilitate estimates.  The norm of $\phi_0$ restricted to a heavy interval is bounded below by 
\be m_i \geq m_i 2\sqrt{L_i}\max(\delta_i^L, \delta_i^R) \geq m_i \sqrt{L_i} \max(\delta_i^L, \delta_i^R)\ee
where the last term is the norm of the constant function $m_i\max(\delta_i^L, \delta_i^R)$ on the interval.  This means that a sine wave with this norm achieves its maximum rather than being nearly flat.\\

{\bf Lemma 1}: The kinetic energy of the ground state restricted to an interval in $M_{heavy}$ is bounded below by
\be m_i^2\left(1 - \frac{1}{\sqrt{2}}\right)^2\frac{\pi^2}{(L_i+1)^2} \ee

{\bf Proof of Lemma 1}:
 The kinetic energy of a heavy interval is bounded below by the kinetic energy of the sine wave with norm $m_i$ and boundary conditions $m_i\delta_i^L$ and $m_i\delta_i^R$.  The energy of the minimizer $\frac{c_im_i}{\sqrt{L_i+1}}\sin(\frac{s_i\pi x}{L_i+1} + t_i)$ is $m_i^2\sin^2(\frac{s_i\pi}{L_i+1})$.  To solve for $s_i$, note that the function must satisfy the boundary conditions 

\be \frac{c_im_i}{\sqrt{L_i+1}}\sin(t_i) = m_i\delta_i^L \ee
\be \frac{c_im_i}{\sqrt{L_i+1}}\sin(s_i\pi + t_i) = m_i\delta_i^R \ee
The left boundary condition is solved using the inverse of sine on $[0, \frac{\pi}{2}]$, $\arcsin(x)$.  The right boundary condition is solved using the inverse of sine on $[\frac{\pi}{2}, \frac{3\pi}{2}]$, $-\arcsin(x) + \pi$.  We obtain as in~\cite{BishopWehr12}:

\be t_i = \arcsin\left(\frac{\delta_i^L\sqrt{L_i+1}}{c_i}\right) \ee
\be s_i\pi + t_i = \pi - \arcsin\left(\frac{\delta_i^R\sqrt{L_i+1}}{c_i}\right)\ee
Solving for $s_i$:

\be s_i = 1 - \frac{1}{\pi}\left(\arcsin\left(\frac{\delta_i^L\sqrt{L_i+1}}{c_i}\right) + \arcsin\left(\frac{\delta_i^R\sqrt{L_i+1}}{c_i}\right)  \right) \ee
Since $\arcsin(\theta) \leq \frac{\pi \theta}{2}$ for $\theta \leq 1$, 

$$ s_i \geq 1 - \frac{\max(\delta_i^L, \delta_i^R)\sqrt{L_i+1}}{c_i} $$
\be  \geq 1 - \frac{1}{\sqrt{2}} \ee
where $c_i \geq 1$, $L_i \geq 1$, and the definition of a heavy interval is used.  \hfill $\Box$ \\

{\bf Lemma 2}:  In the limit as $n \to \infty$, with probability one,
\be \liminf_{n\to\infty} \| \phi_0 |_{M_{heavy}}\|^2 \geq 1 - O\left(\frac{1}{\sqrt{\log_p(g\rho)}}\right) - O\left(\frac{1}{4 b\log_p(g\rho)}\right) \ee
where $O(\cdot)$ is taken with respect to the limit $g\rho \to 0$.

{\bf Proof of Lemma 2}:
Because the potential energy of the ground state must be less than the upper bound on the ground state energy, the norm of the ground state restricted to high potential is bounded as follows: 

\be \|\phi_0|_{M_{b}}\|^2 \leq \frac{C_+}{b \left( \log_p (g\rho)\right)^2} \ee
An upper bound on thehe norm of the ground state on intervals longer than $\log_p(g\rho)$ follows from the fact that the interaction energy must be bounded above by the upper bound on the ground state energy.  The minimum of the interaction energy depends on the number of sites occupied and the norm restricted to the set of these sites.  It was shown in Theorem 1 that for $\|\phi\| = m'$, the minimum of $\sum_{i=1}^L |\phi(i)|^4$ is $\frac{m'^4}{L}$; this result is also due to the inequality between arithmetic and quadratic means.  If $m$ is the norm of $\phi_0|_{M_{long}}$, then the interaction energy for the sites in $M_{long}$ is bounded below, using Theorem 1. 

\be \frac{g\rho L m^4}{\sum_{L_i > \log_p(g\rho)} L_i} \geq \frac{g\rho  m^4 (\frac{n}{pq} - o(n))}{\frac{n g \rho} {pq}(p+pq\log_p(g\rho)) + o(n)} \ee
Using the upper bound on the ground state energy, the left hand side of the above inequality is bounded above by $\frac{C_+}{\left(\log_p(g\rho)\right)^2}$.  In the limit as $n\to \infty$, 

\be \|\phi_0|_{M_{long}}\|^2 \leq \sqrt{\frac{pqC_+}{\log_p(g\rho)} + \frac{pC_+}{(\log_p(g\rho))^2}}\ee
For light intervals, 
\be \frac{m_i^2 }{4L_i} < m_i^2 \max(\delta_i^L,\delta_i^R)^2\ee  
The upper bound on the norm of the ground state restricted to sites of high potential bounds the norm on the boundary points
\be \sum_i m_i^2 \max(\delta_i^L,\delta_i^R)^2 \leq \frac{2C_+}{b \left( \log_p (g\rho)\right)^2} \ee
where the extra factor of $2$ is included to cover the cases where two intervals of zero potential are separated by a single site of positive potential.  The bound on the norm of the ground state restricted to light intervals, which by definition are shorter than $\log_p(g\rho)$, is 

$$\frac{ 1}{\log_p(g\rho)}\sum_{ light\ intervals} m_i^2 \leq  \sum_{light\ intervals}\frac{ m_i^2}{L_i}$$
$$\leq \frac{1}{4}\sum_{light\ intervals} m_i^2\max(\delta_i^L,\delta_i^R)^2 $$
\be \leq \frac{C_+}{4 b \left( \log_p (g\rho)\right)^2} \ee
This means that the norm on light intervals is bounded above by
\be \|\phi_0|_{M_{light}}\|^2 \leq \frac{C_+}{4 b \log_p (g\rho)}\ee
The normalization condition requires
\be \|\phi_0|_{M_b}\|^2 +\|\phi_0|_{M_{long}}\|^2+ \|\phi_0|_{M_{light}}\|^2+\|\phi_0|_{M_{heavy}}\|^2=1\ee
Using the upper bounds on the three other terms, the norm of the ground state restricted to $M_{heavy}$ gives the desired lower bound
$$ \| \phi_0 |_{M_{heavy}}\|^2 \geq 1 - O\left(\frac{1}{\sqrt{\log_p(g\rho)}}\right) - O\left(\frac{1}{ b\log_p(g\rho)}\right) $$
 \hfill $\Box$\\
The energy for a heavy interval is bounded below by
\be \frac{g\rho L m_i^4}{2 L_i} + m_i^2 \left(1 - \frac{1}{\sqrt{2}} \right)^2\frac{\pi^2}{(L_i+1)^2}\ee
where the first term is the minimum of the interaction energy and the second term is the minimum of the kinetic energy.  Using Lagrange multiplier method, the choice of $m_i$, where $i$ labels the heavy intervals, which minimize this lower bound 
\be \sum_i \frac{g\rho L m_i^4}{2 L_i} + m_i^2 \left(1 - \frac{1}{\sqrt{2}} \right)^2\frac{\pi^2}{L_i^2}\ee 
under normalization constraint
\be \sum_i m_i^2 = 1 - O\left(\frac{1}{\sqrt{\log_p(g\rho)}}\right) - O\left(\frac{1}{ b\log_p(g\rho)}\right)\ee
must satisfy
$$ \frac{\partial}{\partial m_i} \left[\sum_i \frac{g\rho L m_i^4}{2 L_i} + m_i^2 \left(1 - \frac{1}{\sqrt{2}} \right)^2\frac{\pi^2}{L_i^2} \right] $$
\be = \lambda \frac{\partial}{\partial m_i} \sum_i m_i^2 \ee
This equation implies that
\be m_i = 0 \ee
or
\be m_i^2 = \frac{L_i}{g\rho L}\left( \lambda - \left(1-\frac{1}{\sqrt{2}}\right)^2\frac{\pi^2}{L_i^2}\right) \ee
For an interval to contribute to the minimization of the lower bound, the kinetic energy of a heavy interval must be less than $\lambda$.  For the kinetic energy to meet this bound, the heavy interval must have length 
\be L_i > \left(1 - \frac{1}{\sqrt{2}} \right)\frac{\pi}{\sqrt{\lambda}} \ee
It follows from Lemma 2 that the normalization condition requires $\lambda$ to satisfy
\be \sum _{L_i > \left(1 - \frac{1}{\sqrt{2}} \right)\frac{\pi}{\sqrt{\lambda}}} \frac{
\lambda L_i}{g\rho L}  =  1 -  O\left(\frac{1}{\sqrt{\log_p(g\rho)}}\right) - O\left(\frac{1}{ b\log_p(g\rho)}\right)\ee
Using equation (29) and approximating $\lfloor x \rfloor$, 
\be \sum _{L_i > x} L_i = \left(\frac{n}{pq} + o(n)\right)\left( \lfloor x\rfloor qp^{\lfloor x\rfloor + 1} + p^{\lfloor x\rfloor+1}\right)\ee
which requires 
\be \frac{\lambda}{g\rho \left(\frac{n}{pq} + o(n)\right) }\left(\frac{n}{pq} + o(n)\right) \left( \left(1 - \frac{1}{\sqrt{2}}\right)\frac{\pi}{\sqrt{\lambda}} qp^{\left(1 - \frac{1}{\sqrt{2}} \right)\frac{\pi}{\sqrt{\lambda}} } + p^{ \left(1 - \frac{1}{\sqrt{2}} \right)\frac{\pi}{\sqrt{\lambda}} }\right) \approx 1 \ee
The asymptotic behavior of the parameter $\lambda$ is determined by $g\rho$.  If $\lambda$ is constant or taken to infinity, the normalization will not hold because the left side of (63) goes to infinity.  If $\lambda$ converges to zero, the dominant term is the exponential.  The correct asymptotic solution for $\lambda$ as $g\rho \to0$ is 
\be \left(1 - \frac{1}{\sqrt{2}} \right)\frac{\pi}{\sqrt{\lambda}} = \log_p(g\rho) + \log_p\left(\log_p(g\rho)\right)  + O(1)\ee
This substitution will make left hand side (63) converge to a constant.  Therefore, the heavy intervals must be longer than $\log_p(g\rho) + \log_p\left(\log_p(g\rho)\right) +O(1)$.  The size of $M_{Heavy}$ is bounded above by the number of sites on intervals longer than this lower bound.  This number is bounded above by 
\be n g\rho \log_p(g\rho)\left(\log_p(g\rho) + \log_p\left(\log_p(g\rho)\right)\right) + o(n) \ee
The interaction energy of any state supported on this number of sites is bounded below by 
\be \frac{g\rho \left(\frac{n}{pq} + o(n) \right)}{2n g\rho \log_p(g\rho)[\log_p(g\rho) + \log_p\left(\log_p(g\rho)\right)] + o(n)} \ee
The lower bound follows. \hfill $\Box$\\

\section{Acknowledgements}
The authors would like acknowledge M. Lewenstein, A. Sanp\'{e}ra, P. Massignan, and J. Stasi\'{n}ska for useful discussions following~\cite{ICFO12} and R. Sims, and L. Friedlander for mathematical insight.  The authors would like to thanks G. Modugno for discussions on experiments related to this work.  \\
The work was partially supported by the NSF grant DMS 0623941.\\
\bibliography{bibmaster}
\bibliographystyle{plain}

\end{document}